\documentclass[conference,10pt]{IEEEtran}
\IEEEoverridecommandlockouts
\usepackage{cite}
\usepackage{amsmath,amssymb,amsfonts}
\usepackage{algorithmic}
\usepackage{graphicx}
\usepackage{mathrsfs}
\usepackage{textcomp}
\usepackage{xcolor}
\def\BibTeX{{\rm B\kern-.05em{\sc i\kern-.025em b}\kern-.08em
		T\kern-.1667em\lower.7ex\hbox{E}\kern-.125emX}}

\usepackage{hyperref}%

\usepackage[caption=false]{subfig}
\usepackage{tikz}
\usetikzlibrary{shapes, arrows, calc, positioning}

\begin{document}
	
	\title{Sensing the Environment with 5G Scattered Signals (5G-CommSense): A Feasibility Analysis}
	
	\author{\IEEEauthorblockN{Sandip Jana}
		\IEEEauthorblockA{\textit{Dept. of Electrical Engineering} \\
			\textit{Indian Institute of Technology}\\
			Hyderabad, India\\
			ee20resch11013@iith.ac.in}
		\and
		\IEEEauthorblockN{Amit Kumar Mishra}
		\IEEEauthorblockA{\textit{Dept. of Electrical Engineering} \\
			\textit{University of Cape Town}\\
			Cape Town, South Africa\\
			akmishra@ieee.org}
		\and
		\IEEEauthorblockN{Mohammed Zafar Ali Khan}
		\IEEEauthorblockA{\textit{Dept. of Electrical Engineering} \\
			\textit{Indian Institute of Technology}\\
			Hyderabad, India\\
			zafar@ee.iith.ac.in}
	}
	
	\maketitle
	
	\begin{abstract}
		By making use of the sensors and AI (SensAI) algorithms for a specialized task, Application Specific INstrumentation (ASIN) framework uses less computational overhead and gives a good performance. This work evaluates the feasibility of the ASIN framework dependent Communication based Sensing (CommSense) system using 5th Generation New Radio (5G NR) infrastructure. Since our proposed system is backed up by 5G NR infra, this system is termed as 5G-CommSense. In this paper, we have used NR channel models specified by the 3rd Generation Partnership Project (3GPP) and added white Gaussian noise (AWGN) to vary the signal to noise ratio at the receiver. Finally, from our simulation result, we conclude that the proposed system is practically feasible.
	\end{abstract}
	
	
	\section{Introduction}
	Inspired by thousands of years of inter-species co-evolution in nature, the CommSense system takes advantage of existing communication infrastructure\footnote{In recent literature \cite{Mauro2021:CIR, Yang2019:CIR} the similar concept is being referred to as Channel Impulse Response (CIR) or Impulse Radio (IR) based sensing.} to sense the environment. CommSense system is based on ASIN framework \cite{A.Mishra2021:ASIN}, which is dedicated to a specific task and might use crude resolution sensors that require less computational power. A CommSense system exploits the reference symbols used in communication and senses the immediate environment, which may not require high resolution, unlike conventional radar, where we compute the precise range and Doppler of a target. In our paper, we are using the upcoming 5G NR telecommunication standard to explore the possibility of sensing using a 5G NR system. To do this, we simulate the sensing results using the ASIN framework for the 5G NR channel models \cite{ChannelModels3GPP}.	
	
	For 4th generation Long Term Evolution (4G LTE), there were six channel models defined by International Telecommunication Union (ITU) \cite{Santu2017:feasibility}, which were similar to TDL models. Whereas for 5G NR, 3GPP defines \cite{ChannelModels3GPP} three different cluster delay line (CDL) channel models (i.e. CDL-A, CDL-B, and CDL-C) for non-line of sight (NLOS) scenarios and two models (CDL-D and CDL-E) for line of sight (LOS) scenarios, these CDL models are useful for Multiple Input Multiple Output (MIMO) channel modeling, which is an integral part of the NR. Till now CDL channels have not been used for sensing applications. For simplified evaluations e.g. non MIMO evaluations \cite{ChannelModels3GPP}, 3GPP defines three tapped delay line (TDL) models (i.e. TDL-A, TDL-B, and TDL-C) for NLOS and two models (i.e. TDL-D and TDL-E) for LOS, but these five TDL channels have more number of ``Taps" as compared to ITU specified LTE channels. In this paper, first, we have generated multiple data points for each of the specified channel models. Then by using channel equalization blocks our proposed system can distinguish each of the five channel models from both the CDL and TDL models. Moreover, we are varying the signal to noise ratio (SNR) to verify the performance of the system. The result we got from the simulation is as expected: each of the clusters generated using different channel models is well distinguishable by visual inspection. As the SNR increases, we get more compact clusters and an increase in inter-cluster separations.
	
	The rest of the paper is arranged as follows: Section \ref{Section:Theory} will give the necessary concepts of a CommSense system, TDL model, and CDL model; then Section \ref{Section:Results} gives the simulation results and discusses the feasibility of the proposed system. Finally, conclusion and future scope is given in Section \ref{Section:conclusion}.
	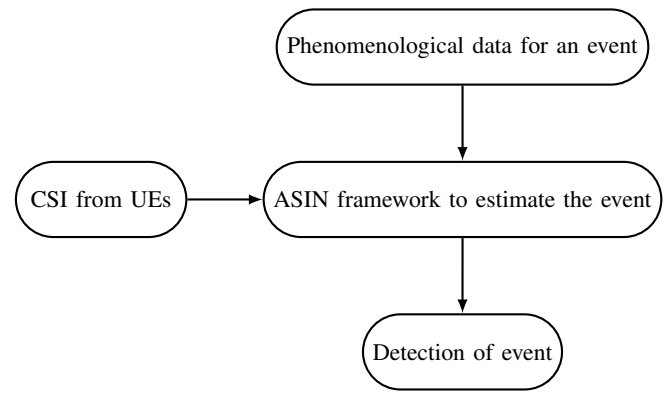
\begin{figure}[b]
		\begin{tikzpicture}[font=\small,thick]
			\node[draw,
			rounded rectangle,
			minimum width=2.5cm,
			minimum height=1cm] (block1) { CSI from UEs };
			
			\node[draw,
			rounded rectangle,
			minimum width=2.5cm,
			minimum height=1cm,
			right=of block1,
			] (block2) { ASIN framework to estimate the event };
			
			\node[draw,
			rounded rectangle,
			minimum width=2.5cm,
			minimum height=1cm,
			above=of block2,
			] (block3) { Phenomenological data for an event };
			
			\node[draw,
			rounded rectangle,
			minimum width=2.5cm,
			minimum height=1cm,
			below=of block2,
			] (block4) { Detection of event };
			
			\draw[-latex] (block1) edge (block2)
			(block3) edge (block2)
			(block2) edge (block4);
			
		\end{tikzpicture}
		\caption{ CommSense flow-diagram to detect an event of interest}
		\label{fig:CommSense_flow_diagram}
	\end{figure}
	\section{Theory}
	\label{Section:Theory}
	
	\subsection{CommSense System}

	\begin{figure*}[!ht]
		\centering
		\includegraphics[scale=0.9]{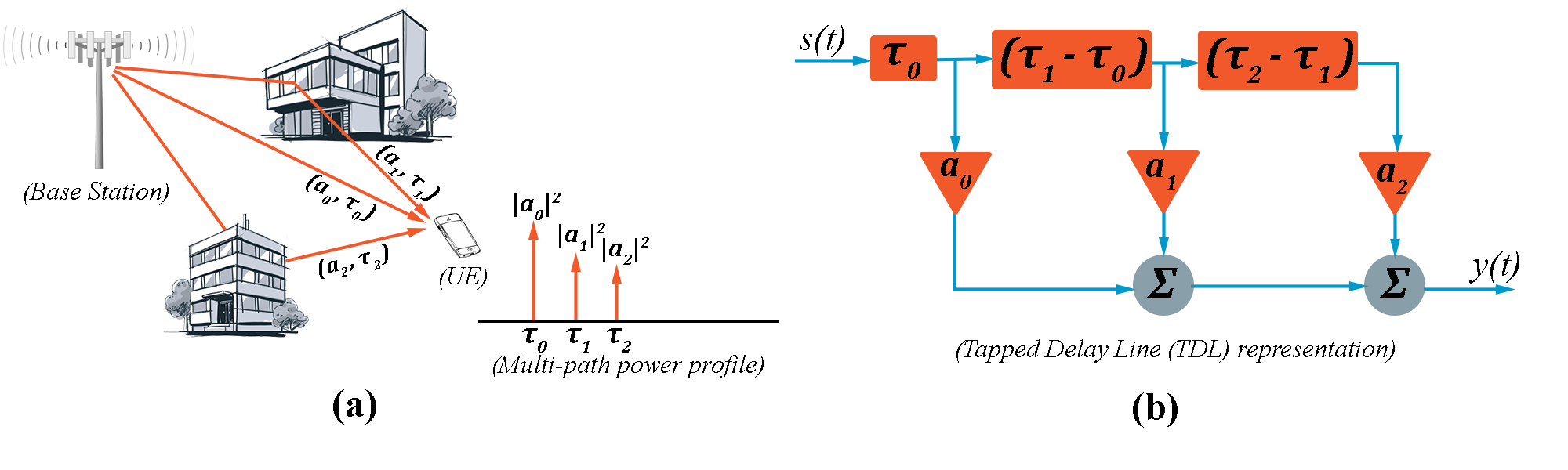}
		\caption{An example Tapped Delay Line (TDL). $a_i$ and $\tau_i$ denotes scaling factor and delay of the signal corresponding to $i^{th}$ path. In (a), we are considering three scattered paths from gNodeB to the user equipment (UE) multipath power profile is given. TDL representation of the given wireless channel is given in (b)}
		\label{fig:TDL_example}
	\end{figure*}
	
	In most communication standards, to filter out the channel modulation, we estimate the fading and non-ergodic wireless channel by sending pilot symbols. As illustrated in Fig.\ref{fig:CommSense_flow_diagram}, we extract the channel state information (CSI) from channel equalization blocks at the receiver or user equipment (UE). In the ASIN framework, first, we gather information about the event of interest, then this stored data and extracted CSI is used for training and pattern classification; then it gives the prediction (i.e. if the event of interest has occurred or not).
    
	\subsection{Tapped Delay Line (TDL)}
    
	A ``Tap'' refers to a point in the delay line that introduces a certain amount of delay and optionally scales the signal \cite{TDL_example}. To model the multi-path nature of the wireless channel shown in Fig.\ref{fig:TDL_example}(a), the signals from each tap are summed. The difference equation of the TDL from Fig.\ref{fig:TDL_example}(b) can be given by,\\
	\begin{equation}
		y(t)=a_0s(t-\tau_0)+
		a_1s(t-\tau_1)+
		a_2s(t-\tau_2)
		\label{eq:TDL_example}
	\end{equation}
	The corresponding transfer function from eq.\ref{eq:TDL_example} is given as,
	\begin{equation}
		H(z) = a_0 z^{-\tau_0} + a_1 z^{-\tau_1} + a_2 z^{-\tau_2}
		\label{eq:TDL_example_TF}
	\end{equation}

	\subsection{Cluster Delay Line (CDL)}
	\begin{figure}[htbp]
		\centering
		\includegraphics[scale=0.75]{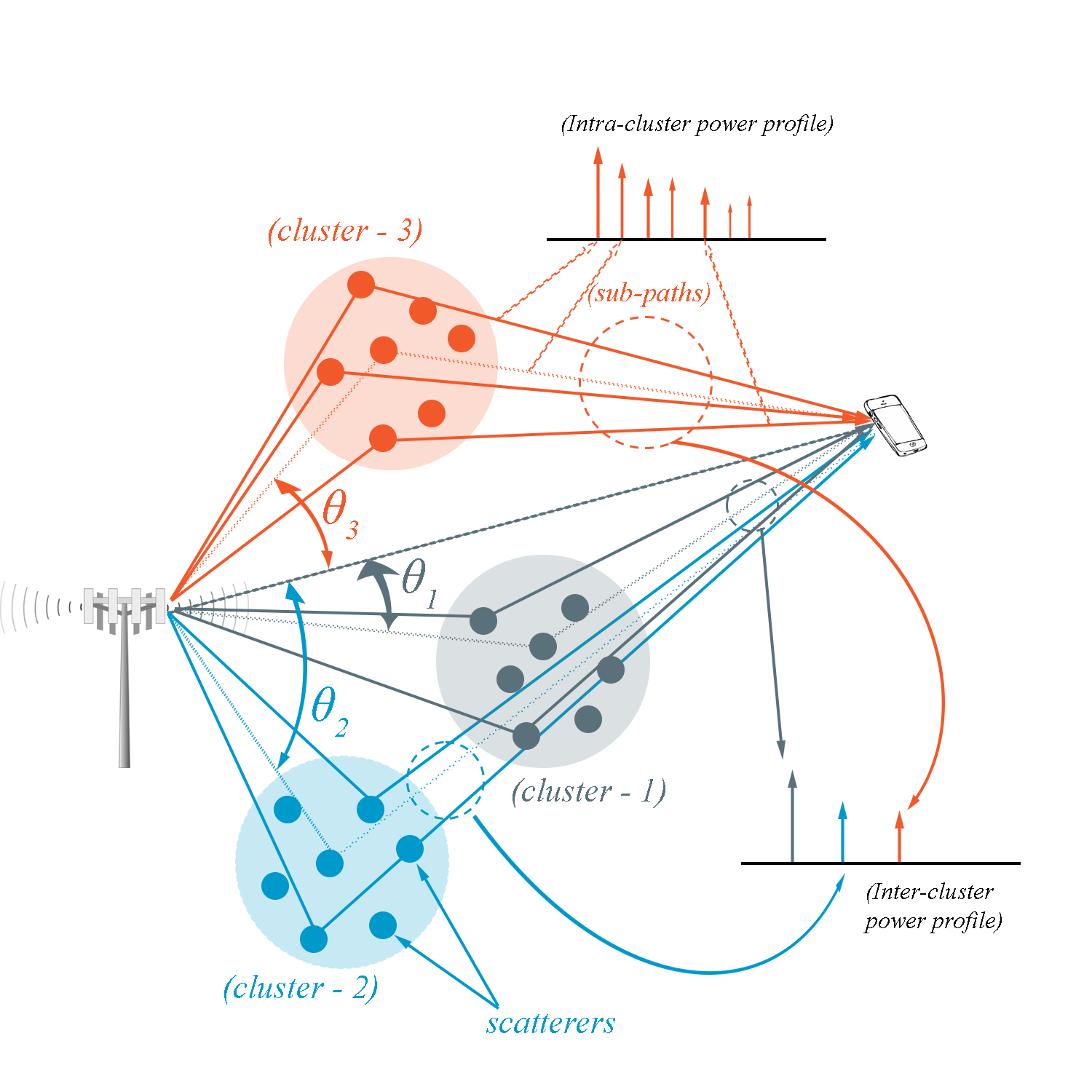}
		\caption{An example of CDL with three clusters; $\theta_i$s represent Angles of Departure (AoD) with respect to the Line of Sight Direction of Departure}
		\label{fig:CDL_example}
	\end{figure}
	
	Sub-paths in wireless channels might get clustered around a particular delay \cite{CDL_example}. A simple illustration of CDL is shown in Fig.\ref{fig:CDL_example}, where three clusters are present and each cluster consists of several scatterers. The power profile for each sub-path and each cluster is shown in terms of the impulse response. 
	
	For our simulation we are using the 3GPP standard \cite{ChannelModels3GPP}; for the TDL channel models normalized delay and scaling factor for each tap, while for the CDL channels normalized delay, power, angle of departure (AoD), angle of arrival (AoA), and other angle measurements for each sub-paths are mentioned in details.

	\section{Simulated Results and Feasibility}
	\label{Section:Results}
	\begin{figure*}[htbp]
		\subfloat[SNR = 0dB]{\begin{minipage}[c][1\width]{0.3\textwidth}
				\centering
				\includegraphics[width=.95\linewidth]{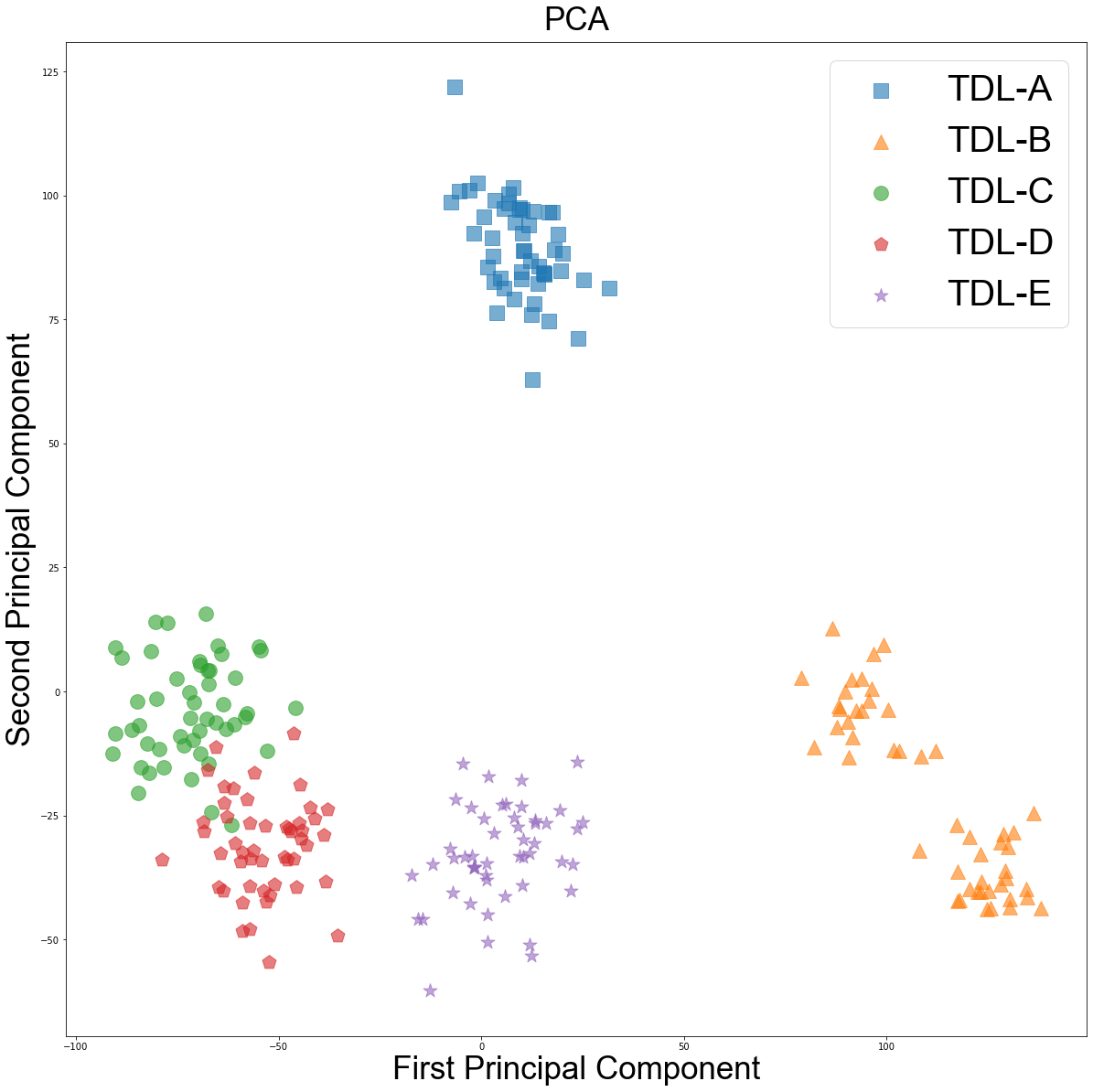}  
				\label{fig:TDL_SNRdB_00}
		\end{minipage}}
		\hfill
		~
		\subfloat[SNR = 10dB]{\begin{minipage}[c][1\width]{0.3\textwidth}
				\centering
				\includegraphics[width=.95\linewidth]{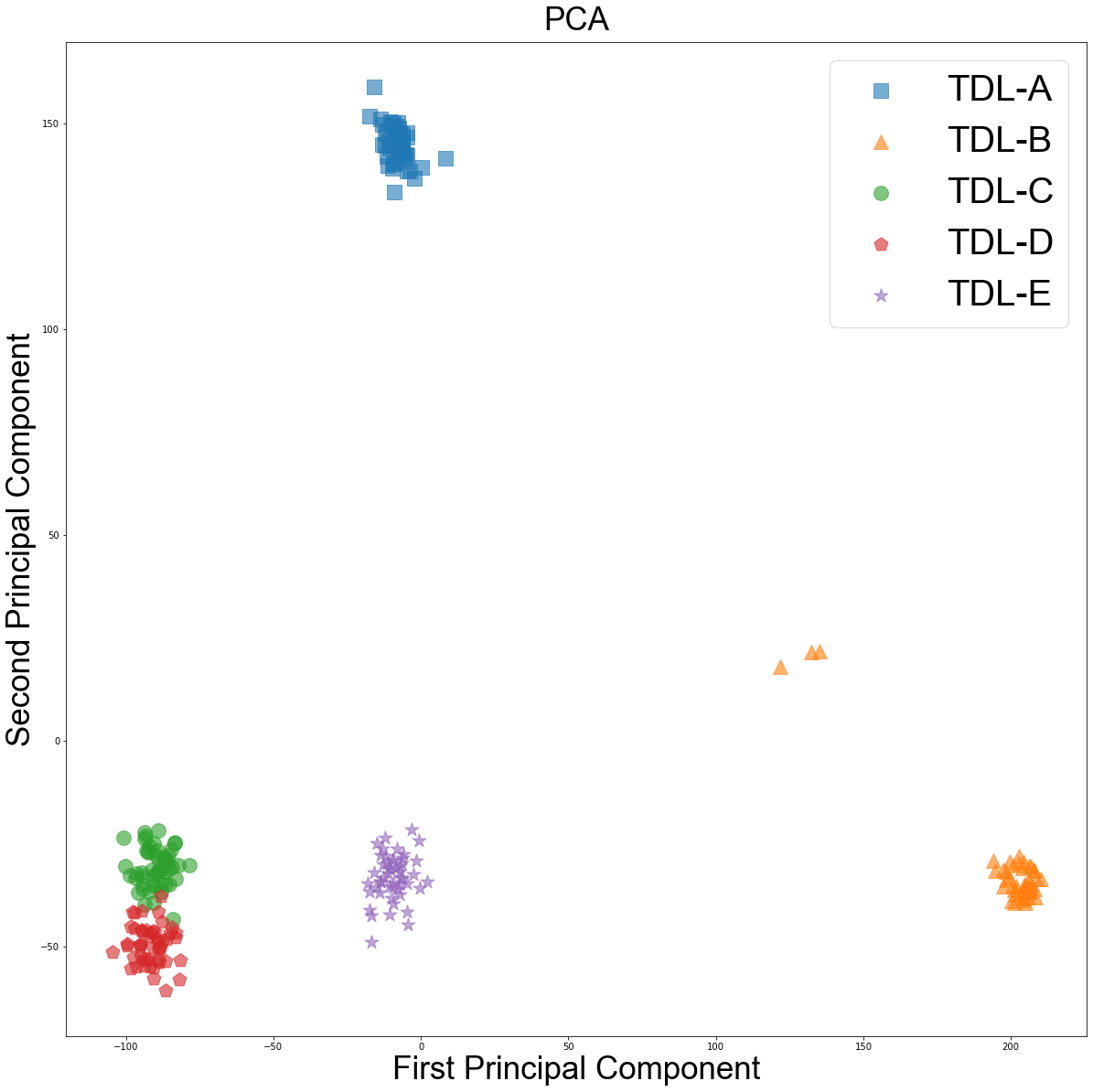}  
				\label{fig:TDL_SNRdB_10}
		\end{minipage}}
		\hfill
		~
		\subfloat[SNR = 20dB]{\begin{minipage}[c][1\width]{0.3\textwidth}
				\centering
				\includegraphics[width=.95\linewidth]{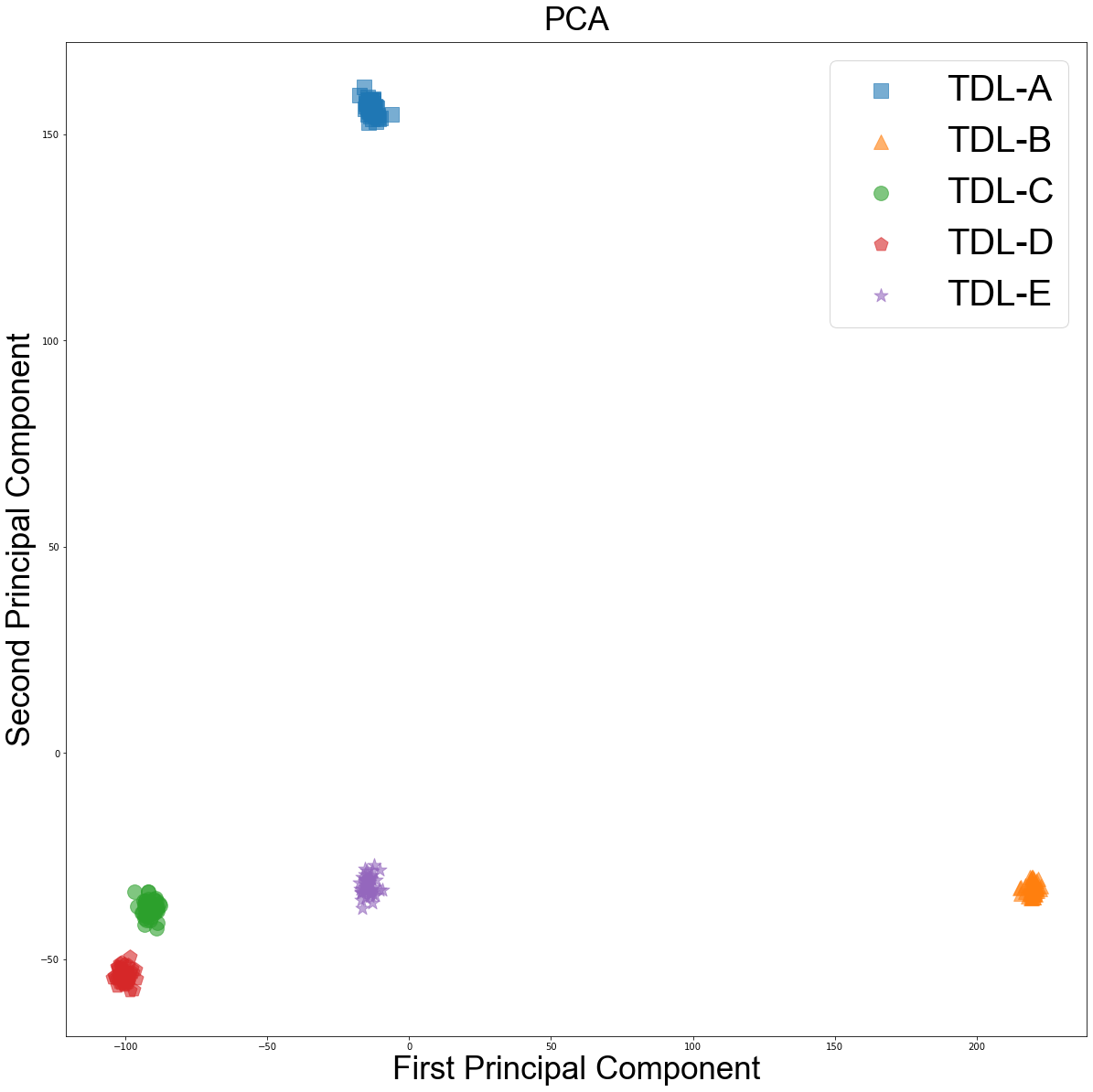}  
				\label{fig:TDL_SNRdB_20}
		\end{minipage}}
		
		\caption{Projection of data (generated using TDL channel models specified by 3GPP) onto the first two Principal Components' direction using PCA and SNR is varied to check the robustness of the proposed system. In (a) For each of the TDL channel model, SNR of each symbol is 0dB, in (b) and (c) SNR is 10dB and 20dB respectively. By using SVM classifier, we get accuracy of $96\%$ for (a), $97.3\%$ for (b) and $100\%$ for (c)}
		\label{fig:TDL_result}
	\end{figure*}
	
	\begin{figure*}[htbp]
		\subfloat[SNR = 0dB]{\begin{minipage}[c][1\width]{0.3\textwidth}
				\centering
				\includegraphics[width=.95\linewidth]{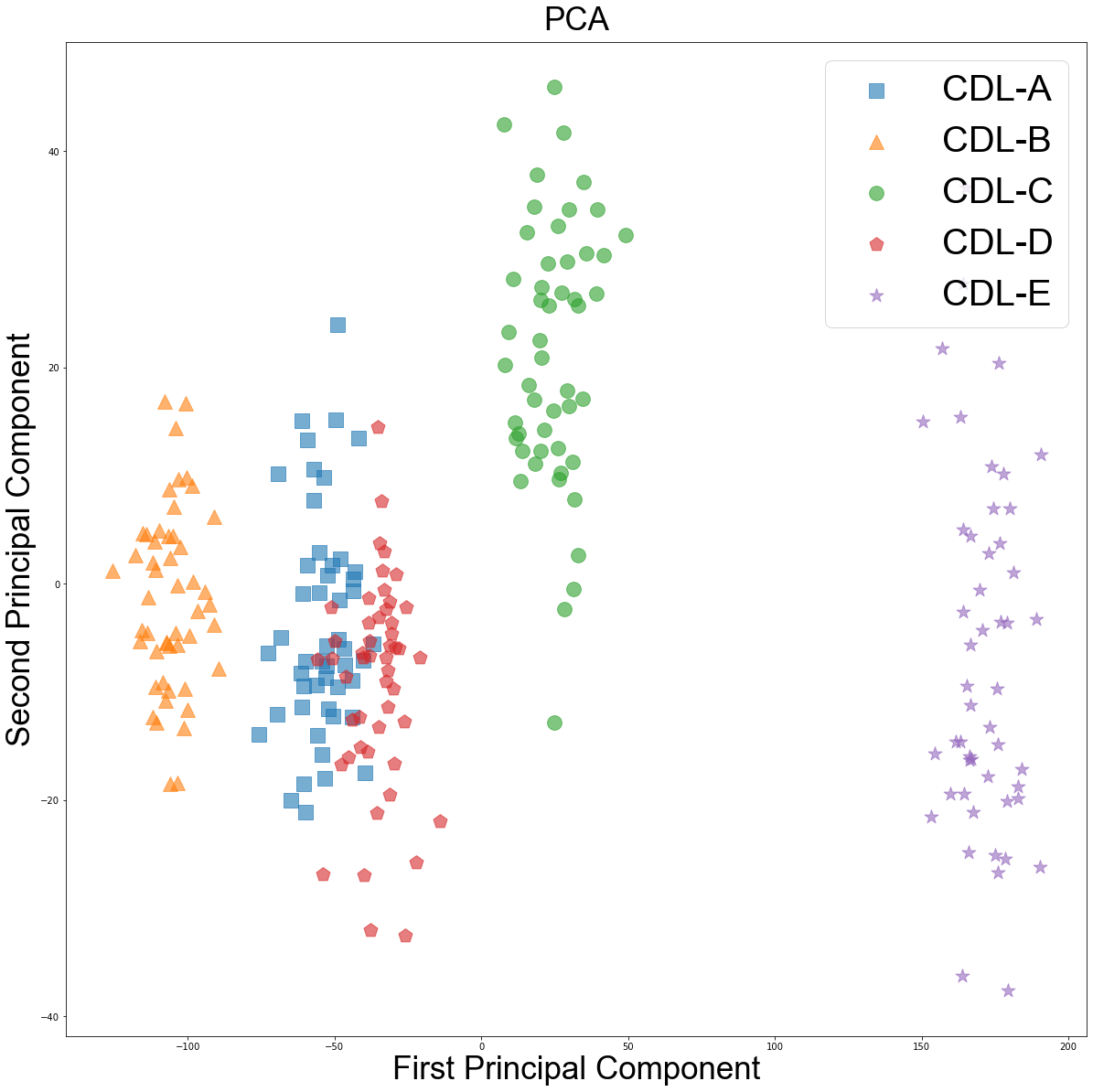}  
				\label{fig:CDL_SNRdB_00}
		\end{minipage}}
		\hfill
		~
		\subfloat[SNR = 10dB]{\begin{minipage}[c][1\width]{0.3\textwidth}
				\centering
				\includegraphics[width=.95\linewidth]{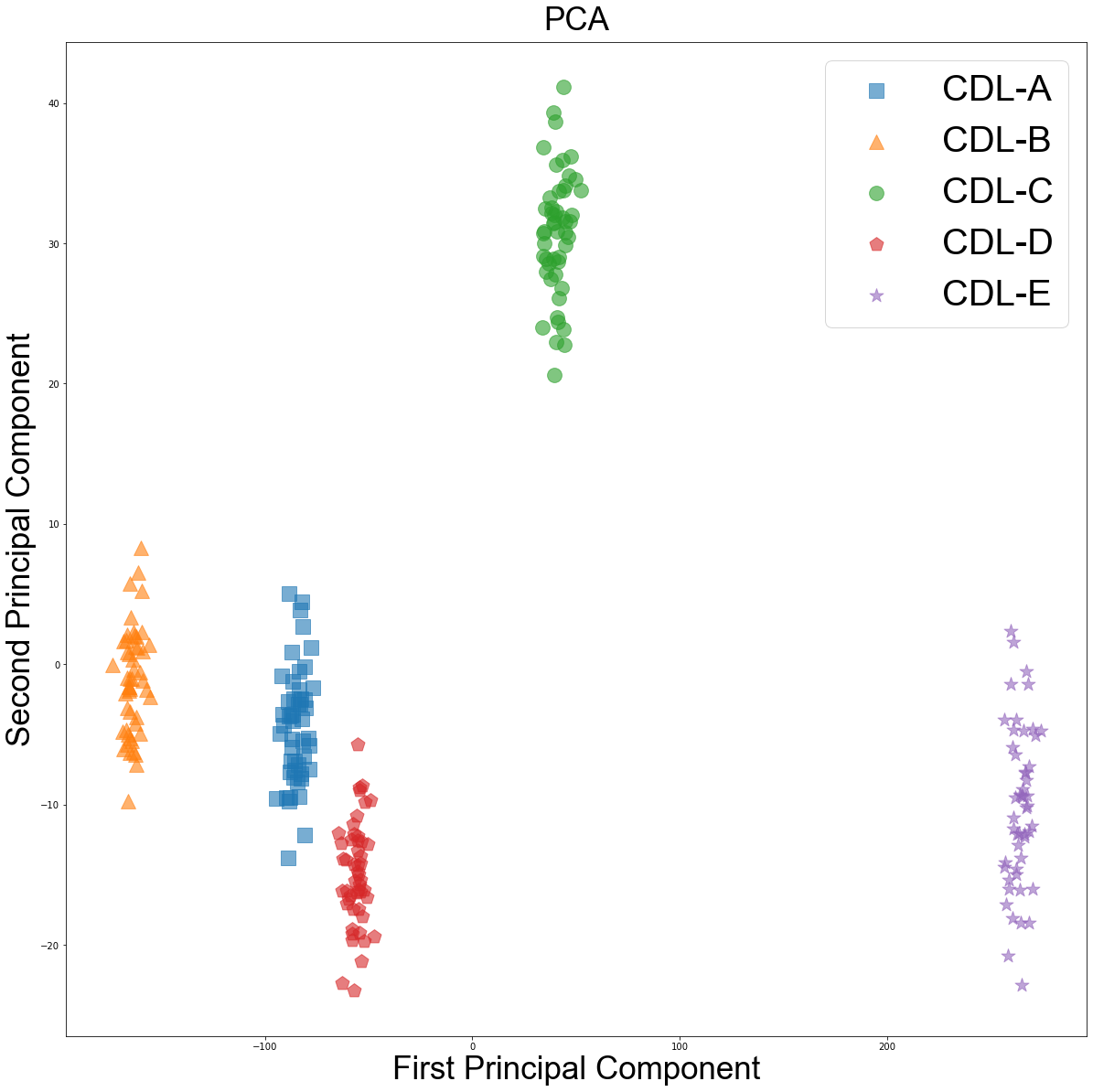}  
				\label{fig:CDL_SNRdB_10}
		\end{minipage}}
		\hfill
		~
		\subfloat[SNR = 20dB]{\begin{minipage}[c][1\width]{0.3\textwidth}
				\centering
				\includegraphics[width=.95\linewidth]{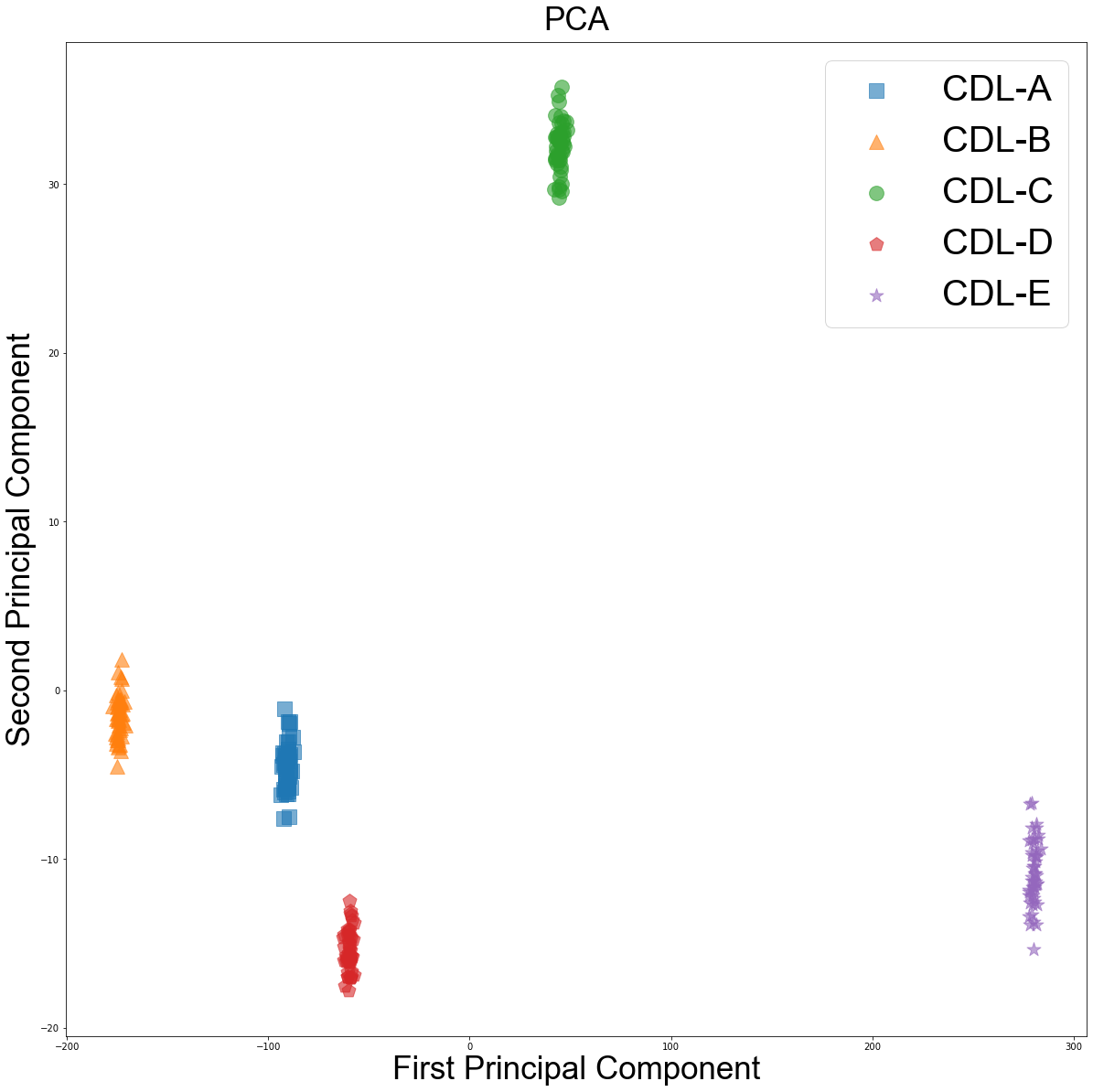}  
				\label{fig:CDL_SNRdB_20}
		\end{minipage}}
		
		\caption{Projection of CDL data onto first two Principal Components' direction using PCA, In (a) For each of the CDL channel models, SNR of each symbol is increased from (a) to (c). By using SVM classifier, we get sensing accuracy of $93.33\%$ for (a), $100\%$ for (b) and $100\%$ for (c)}
		\label{fig:CDL_result}
	\end{figure*}
	
	At the carrier frequency of $4GHz$, for both the TDL and CDL model we generated 50 samples using MATLAB for each of the five channel models according to the 3GPP specifications; then we added AWGN noise to vary the SNR at the receiver; we are considering SNR of 0dB, 10dB and 20dB. After the pilot symbol is received at the UE, we use MMSE based equalizer to estimate the channel. This high-dimensional CSI is projected to two dimensions using PCA \cite{Sandip2022:evaluation}, then we are using the SVM classifier to evaluate the accuracy.
	
	Fig.\ref{fig:TDL_result} shows the results by varying the SNR for five TDL models and accuracy using SVM based classifier; it achieves the accuracy of 96\%, 97.3\% and 100\% at the SNR of 0dB, 10dB and 20dB respectively. Fig.\ref{fig:CDL_result} shows the results for CDL models; here we achieve an accuracy of 93.33\%, 100\% and 100\% for the SNR of 0dB, 10dB and 20dB respectively. In lower SNR, data points in CDL clusters are more dispersed compared to TDL due to the presence of correlated multipaths, but as the SNR increases the classification accuracy goes up as expected.

	\section{Conclusion and Future Work}
	\label{Section:conclusion}
	Our work analyzed the feasibility of environment sensing using 5G NR infrastructure, especially for the CDL channels which was not done before. The performance from our simulation shows that our proposed system is feasible using 5G NR. Since this system can co-exist with the communication system, it will be a cost-effective solution for sensing applications. Here, we have considered one UE in the simulations. In the future study, we will consider multiple UEs and develop an optimal decision fusion algorithm.
	
	\section*{Acknowledgment}
	This work was mainly supported by University Grants Commission (UGC) and partly by project VAJRA, Department of Science and Technology (DST), Govt. of India.
	
	\bibliographystyle{IEEEtran}

\end{document}